\documentclass[preprint2]{aastex63}

\usepackage{natbib}

\usepackage{amsmath}
\usepackage{hyperref}
\usepackage{color}
\usepackage{float}
\usepackage{xspace} 

\newcommand{\angstrom}{\mbox{\normalfont\AA}\xspace}
\newcommand{\kepler}{\textsl{Kepler}\xspace}

\newcommand{\teff}{\ensuremath{T_{\mbox{\scriptsize eff}}}\xspace}
\newcommand{\tspot}{\ensuremath{T_{\mbox{\scriptsize spot}}}\xspace}

\begin{document}

\title{Hunt for Starspots in HARPS Spectra of G and K Stars}

\author[0000-0003-2528-3409]{Brett M.~Morris}
\affiliation{Center for Space and Habitability, University of Bern, Gesellschaftsstrasse 6, 3012 Bern, Switzerland}
 
\author[0000-0001-8981-6759]{H.~Jens Hoeijmakers}
\affiliation{Center for Space and Habitability, University of Bern, Gesellschaftsstrasse 6, 3012 Bern, Switzerland}
\affiliation{Observatoire de Gen\`eve, University of Geneva, Chemin des Maillettes, 1290 Sauverny, Switzerland}

\author[0000-0003-4269-3311]{Daniel Kitzmann}

\author[0000-0002-9355-5165]{Brice-Olivier Demory}

\affiliation{Center for Space and Habitability, University of Bern, Gesellschaftsstrasse 6, 3012 Bern, Switzerland}

\begin{abstract}
We present a method for detecting starspots on cool stars using the cross-correlation function (CCF) of high resolution molecular spectral templates applied to archival high-resolution spectra of G and K stars observed with HARPS/HARPS-N. We report non-detections of starspots on the Sun even when the Sun was spotted, the solar twin 18 Scorpii, and the very spotted Sun-like star HAT-P-11, suggesting that Sun-like starspot distributions will be invisible to the CCF technique, and should not produce molecular absorption signals which might be confused for signatures of exoplanet atmospheres. We detect strong TiO absorption in the T Tauri K-dwarfs LkCa 4 and AA Tau, consistent with significant coverage by cool regions. We show that despite the non-detections, the technique is sensitive to relatively small spot coverages on M dwarfs and large starspot areas on Sun-like stars.
\vspace{2cm}
\end{abstract}

\section{Introduction}
Molecular band modeling (MBM) is a technique for measuring starspot coverage and temperatures \citep[see e.g.:][]{Neff1995, oneal1996, oneal1998, ONeal2001, ONeal2004, Morris2019b}. The MBM technique seeks to describe spectra of active stars as linear combinations of warm (photospheric) and cool (starspot) stellar spectrum components. One useful tracer molecule is TiO, which forms at temperatures $<4000$ K, so cool starspots near or below this temperature will feature TiO absorption, while the rest of the photosphere of a G or K star will not \citep{Vogt1979, Ramsey1980}.

Molecular band modeling is challenging for several reasons. The size of the expected signal generated by TiO absorption is exceptionally small, because Sun-like stars are typically $<1\%$ spotted and the $<1\%$ of the photosphere that is covered in spots is intrinsically dimmer than the rest of the photosphere by $\gtrsim 30\%$ \citep{Solanki2003}. In addition, weak TiO absorption features can be degenerate with continuum normalization over the small wavelength ranges where the TiO absorption is greatest. In addition, all inferences from MBM stem from relying on inaccurate models, for example, due to incomplete linelists. Thus at modest spectral resolutions and signal-to-noise ratios, only the most spotted stars will generate sufficient TiO absorption to be detected confidently with MBM \citep[see discussion in][]{Morris2019b}. 

However, the need for constraints on the starspot covering fractions of planet-hosting stars continues to grow as we seek to observe the transmission and day-side spectra of Earth-sized exoplanets \citep[see e.g.:][]{Rackham2018, Ducrot2018, Morris2018f, Wakeford2019}. The spectral features generated by exoplanet atmospheres may be degenerate with the signatures of starspots, which also vary in time and wavelength primarily because of the stellar rotation. Therefore, the presence of starspots potentially hinders the interpretation of exoplanet observations. If we seek to measure starspot coverages with sufficient precision to mitigate the effects of starspots we likely need to move beyond MBM.

An analogous contrast-ratio problem occurs when detecting the emission or transmission spectra from exoplanet atmospheres, which has been addressed using the cross-correlation (CCF) technique applied to high resolution spectroscopy \citep{Snellen2010,Brogi2012}. The CCF is sensitive to both strong and weak absorption features that occur at {\it all} wavelengths throughout the spectrum of the planet, constructively co-adding the absorption lines when a template spectrum is matched with the observed spectrum at the correct Doppler velocity.

In this work, we seek to use high resolution ($R \sim 115,000$) spectra from the High Accuracy Radial Velocity Planet Searcher \citep[HARPS;][]{Mayor2003} and HARPS-North (HARPS-N) of bright, nearby stars to measure their spot coverages via the cross-correlation function. In Section~\ref{sec:template} we construct template spectra for TiO, CO and H$_2$O.  In Section~\ref{sec:sims} we present simulated observations of spotted stars and examine the significance of spot detections with the cross-correlation technique. In Section~\ref{sec:control} we examine which portions of HARPS spectra are sensitive to each molecule given imperfect line lists, and search for TiO absorption in the spectra of the Sun, 18 Scorpii, HAT-P-11 and two T Tauri stars. We briefly conclude in Section~\ref{sec:conclusion}.

\section{Template Construction} \label{sec:template}

We use the library of stellar atmospheric structures published by \citet{Husser2013A&A...553A...6H}. The structures are based on calculations with the state-of-the-art PHOENIX stellar atmosphere model \citep{Hauschildt1999JCoAM.109...41H}. For this work, we extract the basic structures (temperature-pressure profiles) from this library for a surface gravity of $\log g = 4.50$ and solar elemental abundances. To approximate the atmospheric conditions within the starspots, we use structures for different stellar effective temperatures from 4000 K to 2500 K. 

Based on the temperature-pressure profiles from the PHOENIX library, we calculate theoretical template emission spectra using our \texttt{Helios-o} spectrum calculator. For emission spectra, \texttt{Helios-o} employs a general discrete ordinate radiative transfer model based on the \texttt{CDISORT} code \citep{Hamre2013AIPC.1531..923H}. We use eight streams to compute the emission spectra in this study.

To determine the chemical composition we employ the fast equilibrium chemistry code \texttt{FastChem} published by \citet{Stock2018MNRAS.479..865S}. The solar elemental abundances are taken from \citet{Asplund2009ARA&A..47..481A}.

Collision-induced continuum absorption of He-H, H$_2$-H$_2$, and H$_2$-He pairs is included by using data from the HITRAN database \citep{Karman2019Icar..328..160K}. The description for the continuum cross-section of H$^-$ is taken from \citet{John1988A&A...193..189J}.

To generate the templates for TiO, we use absorption coefficients based on line lists from the Exomol database \citep{McKemmish2019MNRAS.488.2836M}. To generate the template for water, we use absorption coefficients from \citet{Barber2006}.

The theoretical, high-resolution spectra are calculated within the spectral range of HARPS with a constant step size of 0.03 cm$^{-1}$ in wavenumber space. 

\section{Simulated Observations} \label{sec:sims}

\subsection{Definition of the CCF} \label{sec:ccf_def}

We define the cross-correlation function (CCF) for an observed spectrum $x$, given a template spectrum evaluated at a specific velocity $T(v)$,
\begin{equation}
    \mathrm{CCF} = \sum_i x_i T_i(v),
\end{equation}
where we have normalized the template such that it is positive in molecular absorption features and near-zero in the continuum, and
\begin{equation}
    \sum_i T_i(v) = 1.
\end{equation}
This definition of the CCF can be interpreted as a mean of the flux in each echelle order weighted by the values of the spectral template. When the velocity $v$ is incorrect and/or the template does not match the observed spectrum, the weighted-mean flux is near unity (continuum). When the velocity is correct and the template matches absorption features in the observed spectrum, the absorption features in the spectrum ``align'' with the inverse absorption features in the template, and the weighted-mean flux is less than one.  We consider detections of molecules with the CCF to be ``significant'' if the CCF decrement at the correct radial velocity of the star is less than a few standard deviations smaller than the CCF continuum. In this way, the CCF yields a ``mean absorption line'' due to the molecule specified by the template at the velocity of the star.

We provide an open source Python package called \texttt{hipparchus}. The software and documentation are available online\footnote{\url{https://github.com/bmorris3/hipparchus}}.

\subsection{Simulations}

We investigate whether one should expect significant detections of starspots with cross-correlation functions of high resolution spectra by assembling a grid of simulated spectra. Each simulated spectrum contains 4000 wavelength bins, similar to a single echelle order of HARPS. We imagine a star which has uniform continuum emission from its photosphere, with no confounding absorption features. We give the star a spot covering fraction $f_S$, and assign the spotted regions the absorption spectrum of a pure TiO atmosphere.

We simulate noisy spectra of spotted stars by: (1) combining the flux-weighted spectral template with a uniform continuum, given that the flux ratio of the spotted regions compared with the total spectrum will be 
\begin{equation}
    \mathcal{F}_\lambda = \frac{f_S B_\lambda(T_\mathrm{spot})}{(1-f_S) B_\lambda(T_\mathrm{eff}) + f_S B_\lambda(T_\mathrm{spot})},
\end{equation}
given a range of spot coverages from $0.1-50$\%; (2) adding random normal noise to the spotted spectra with signal-to-noise ratio ranging from $10-10^5$, representing low S/N spectra through very deeply stacked spectra; (3) taking the cross-correlation function of the spectral template with the ``observed'' noisy, spotted spectra; and (4) computing the amplitude of the CCF peak in relation to the scatter about the continuum. 

\begin{figure}
    \centering
    \includegraphics[scale=0.85]{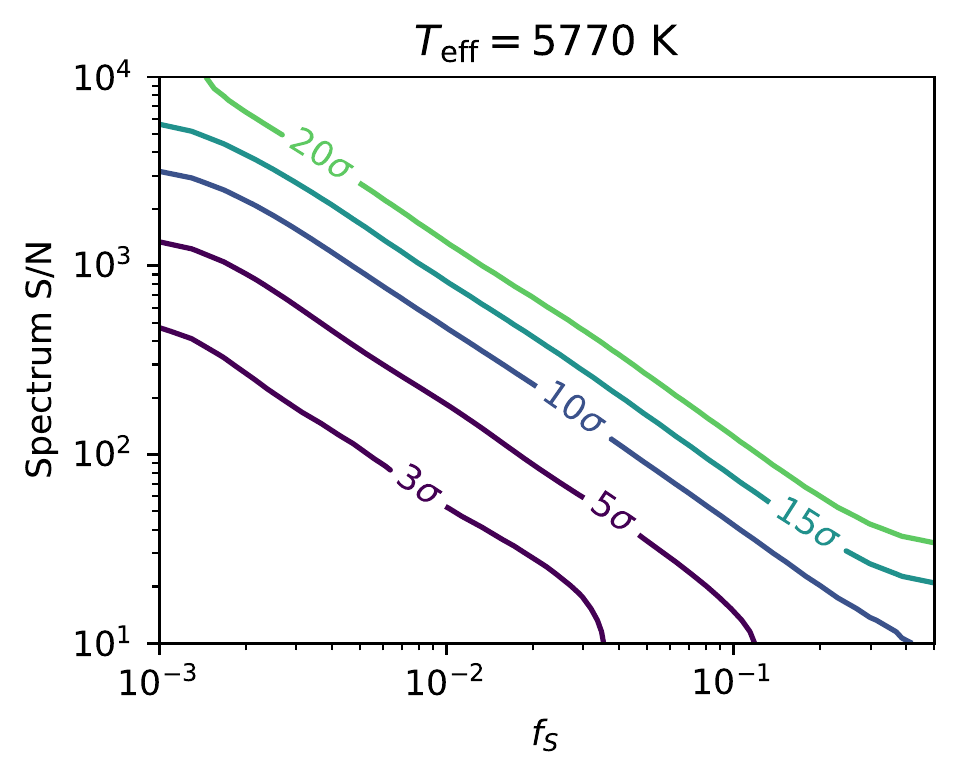}
    \includegraphics[scale=0.85]{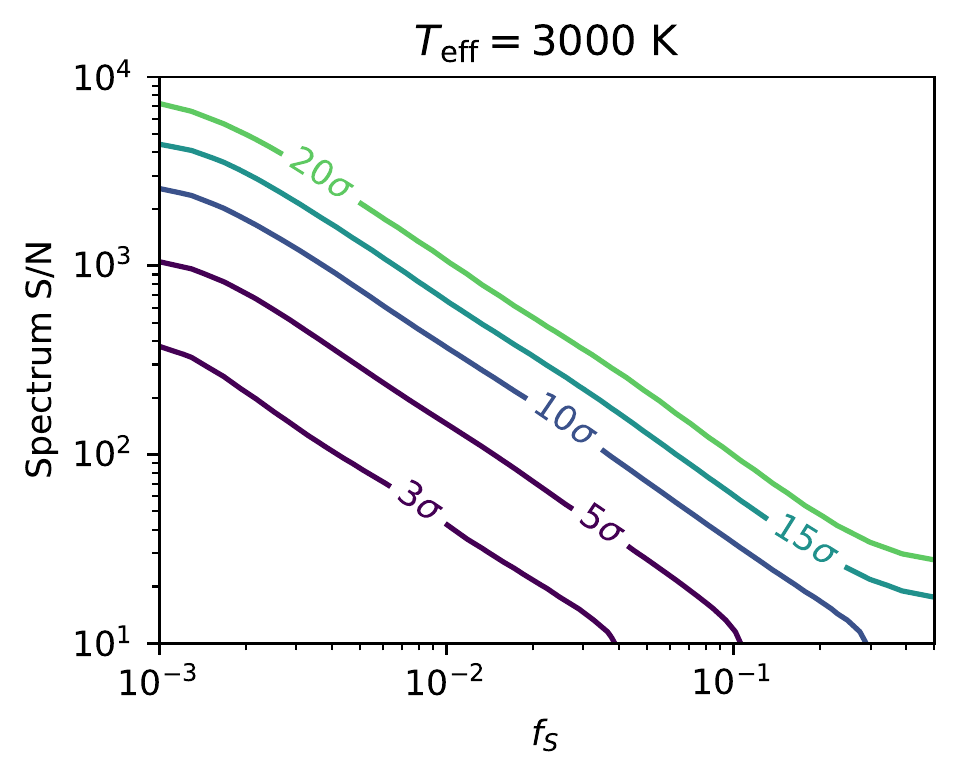}
    \caption{Signal-to-noise ratio of the cross-correlation function (colored contours) as a function of the signal-to-noise ratio of the simulated spectrum, and the spot coverage $f_S$ in the simulated spectrum, for a Sun-like star with $\tspot = 3000$ K (upper) and for an M dwarf with $\tspot = 2500$ K (lower). A typical, high-quality HARPS spectrum of a bright Sun-like star has $S/N \approx 300$, and Proxima Centauri typically has $S/N\approx 200$.}
    \label{fig:snr}
\end{figure}

We plot the S/N curves for the observed CCF as a function of the spot coverage and each spectrum's S/N in Figure~\ref{fig:snr}. Each contour represents the signal-to-noise ratio of the peak in the the cross-correlation function for a given combination of the stellar spectral S/N and spot coverage $f_S$. The upper plot shows the results for a Sun-like star with $\teff = 5770$ K and $\tspot = 3000$ K, and the lower plot represents an M star with $\teff = 3000$ K and $\tspot = 2500$ K. 

We focus first on the Sun-like case in the upper panel of Figure~\ref{fig:snr}. Note that for a typical HARPS spectrum of a bright star with $S/N \sim 300$, and a Sun-like spot coverage $f_S < 5 \times 10^{-3}$, the CCF peak has $S/N < 3$. In other words, Sun-like spot coverages on Sun-like stars should be undetectable with the cross-correlation function in individual exposures. %, confirming our non-detection in Section~\ref{sec:sun}. 
If we imagine a bright Sun-like star with 10\% spot coverage and HARPS spectra with $S/N \sim 300$, the CCF technique is expected to detect starspots at $10\sigma$ confidence. 

The CCF signal is more significant as one inspects stars with smaller \teff and less extreme (warmer) spot temperatures. For the M dwarf in the lower panel of Figure~\ref{fig:snr}, which has a similar effective temperature to Proxima Centauri and $\tspot = 2500$ K, spots are more readily detectable via the CCF. Proxima Centauri is routinely observed by HARPS with $S/N \sim 200$, so watery spot coverages as small as $f_S=0.01$ could be detectable at $5\sigma$ if they were present. 

\section{Hunting for Spots} \label{sec:control}

\begin{figure}
    \centering
    \includegraphics[scale=0.8]{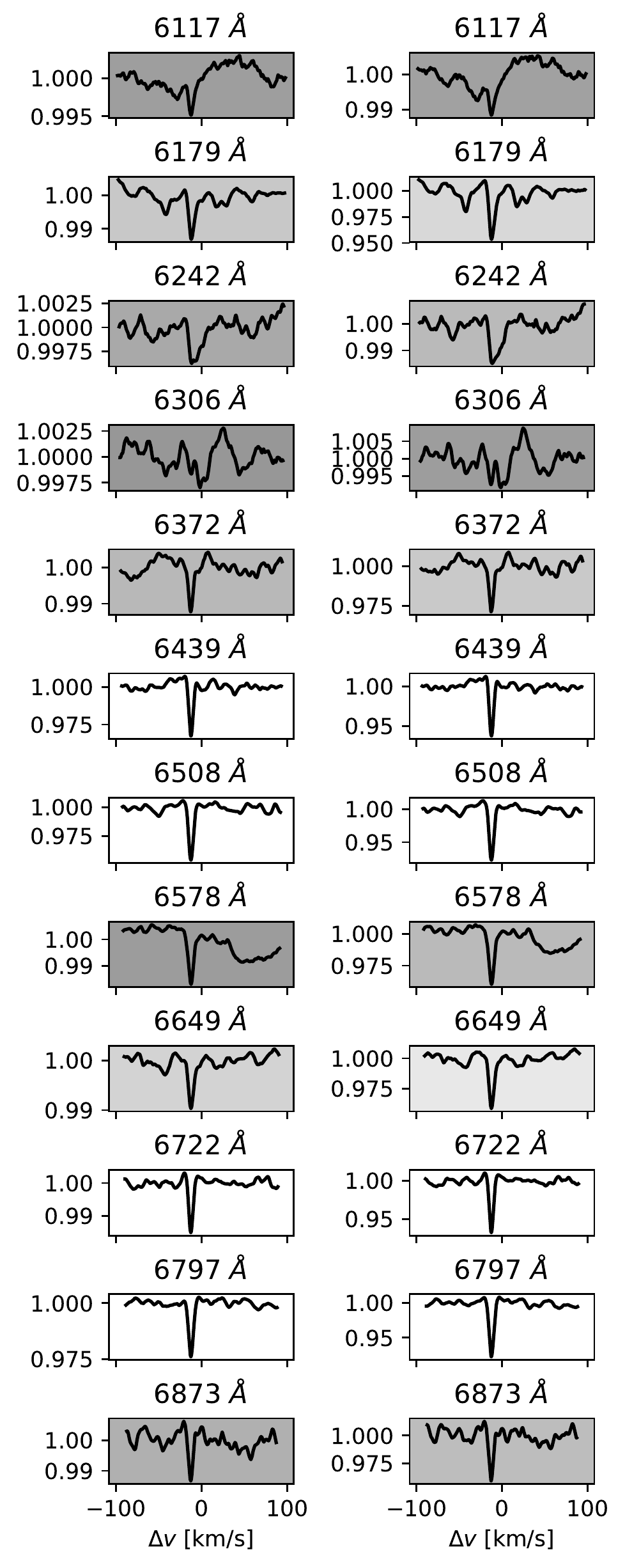}
    \caption{Identifying echelle orders with high-confidence detections of TiO in the stellar spectrum of Proxima Centauri, an M5V star, with a $\teff = 3000$ K (left) and $4000$ K (right).  The background of each plot is shaded proportionately to the S/N ratio of the detection of TiO in that echelle order (dark is low signal, bright is high signal).}
    \label{fig:control}
\end{figure}

\subsection{Testing the TiO line list}

Figure~\ref{fig:control} shows the CCF between an observed spectrum of Proxima Centauri (HARPS Program ID: 072.C-0488(E), PI: M.~Mayor) and the TiO model spectra at 3000 K (left column, matching Proxima Centauri which has $\teff \sim 3000$ K) and 4000 K (right column). Each panel represents one HARPS echelle order, with the central wavelength noted in the title. In black we plot the ``weighted-mean absorption profile'' CCF parameterization outlined in Section~\ref{sec:ccf_def}.

TiO is detected with $S/N > 3$ in most echelle orders, peaking at $S/N \sim 22$ for the TiO template with the correct effective temperature. The order of magnitude variation in the $S/N$ for Proxima Centauri, which clearly has significant TiO absorption in every echelle order redward of 4500 \angstrom (see Figure~\ref{fig:molecules}, in the Appendix), demonstrates that the line list is imperfect, in agreement with \citet{Hoeijmakers2015}.

\subsection{Search for cool spots on Proxima Centauri}
      
In addition to using Proxima Centauri as a control target to identify the orders where the TiO line list is the most reliable, we can search for cool spots on Proxima Centauri. Proxima has $\teff = 2980 \pm 80$ K \citep{Ribas2017}, so we assume dark star-spots may have temperatures of $\sim 2500$ K \citep{Berdyugina2005}, and search for emission from these spots using the cross-correlation function of the spectrum with a template for water at 2500 K. After all, water has even been detected in sunspots \citep{Wallace1995}.

\begin{figure}
    \centering
    \includegraphics[scale=0.7]{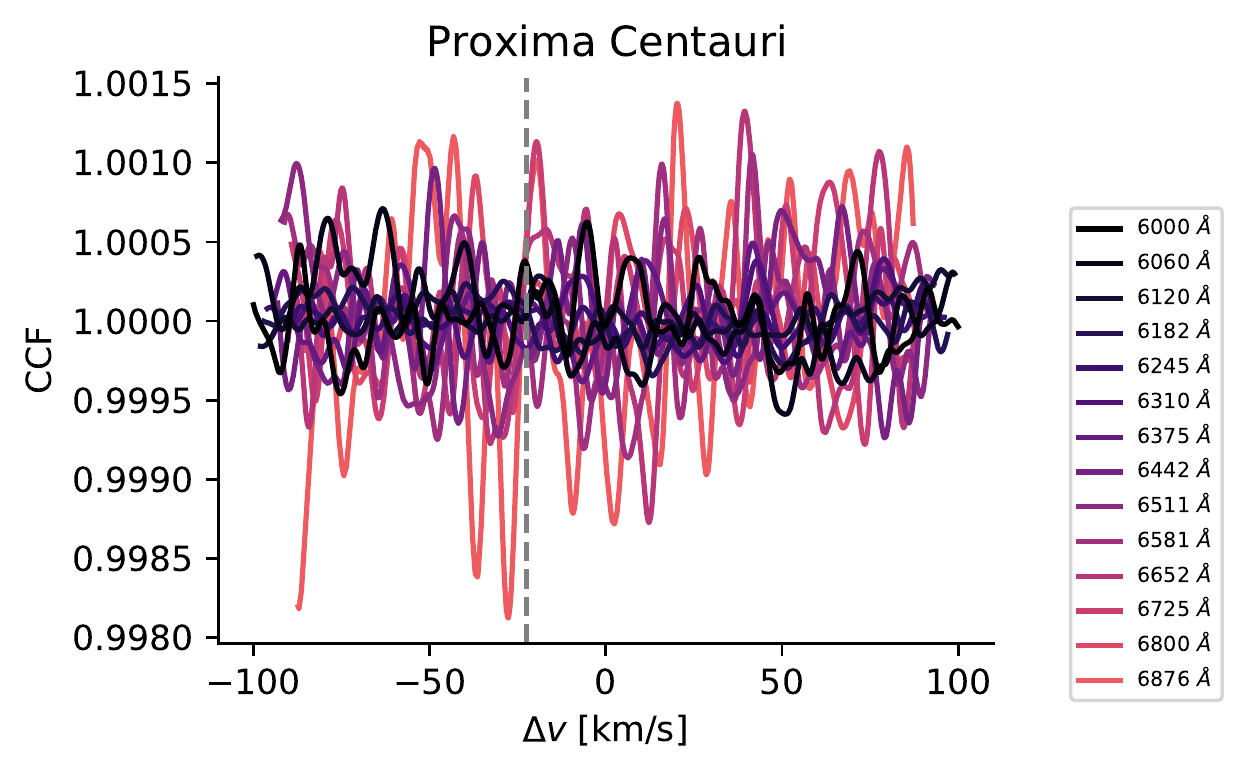}
    \caption{Null detection of water absorption in the high resolution spectrum of Proxima Centauri. For reference, the radial velocity of Proxima is $-22$ km s$^{-1}$, marked with the vertical dashed line. }
    \label{fig:proxcen_water}
\end{figure}

Figure~\ref{fig:proxcen_water} shows that no significant detections of water  absorption are present in any of the spectral orders red-ward of 6000 \AA, where the water spectrum has small absorption features (see Figure~\ref{fig:molecules}). The lack of detectable spots could indicate that there are few cool spots, or their temperature contrast is significantly different from $\Delta T = 500$ K, or the S/N of these spectra are insufficient to detect the relatively weak absorption lines from water.

\subsection{Search for sunspots} \label{sec:sun}

The umbrae of sunspots reach temperatures as low as $\sim$4000 K \citep{Solanki2003}, and therefore we might expect a very spotted Sun to generate TiO absorption. Fortunately, there are also several Sun-observing spacecraft which have been imaging the Sun for decades, often simultaneously with HARPS observations of reflected sunlight via observations solar system targets, such as the Moon. 

\begin{figure*}
    \centering
    \includegraphics[scale=0.85]{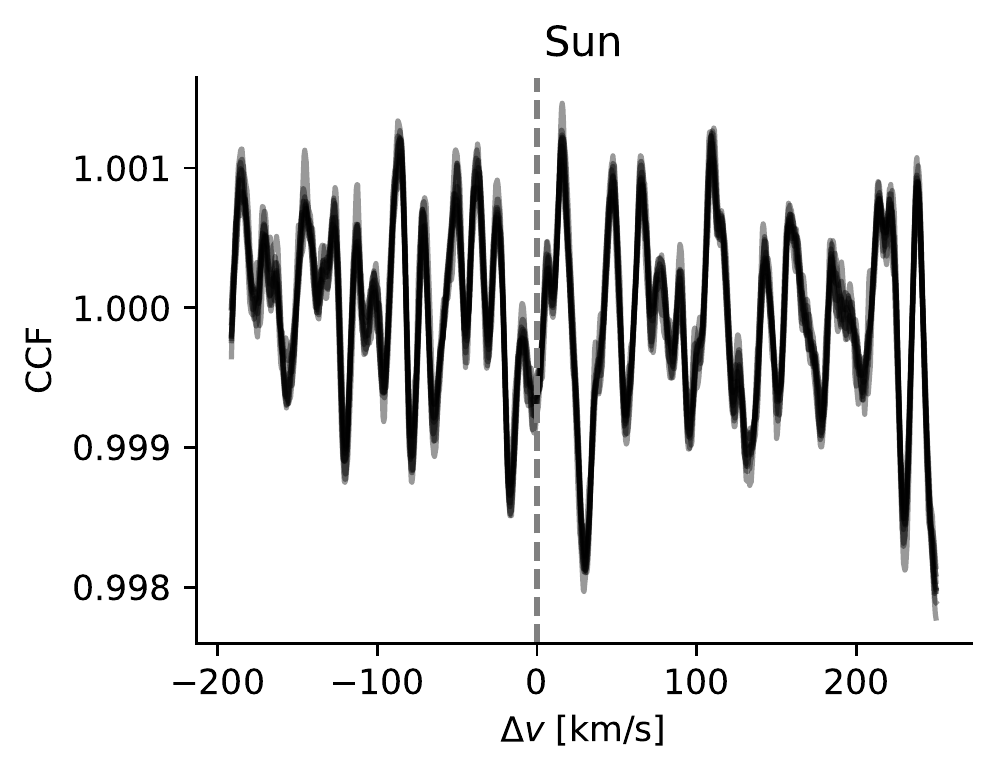}
    \includegraphics[scale=0.7]{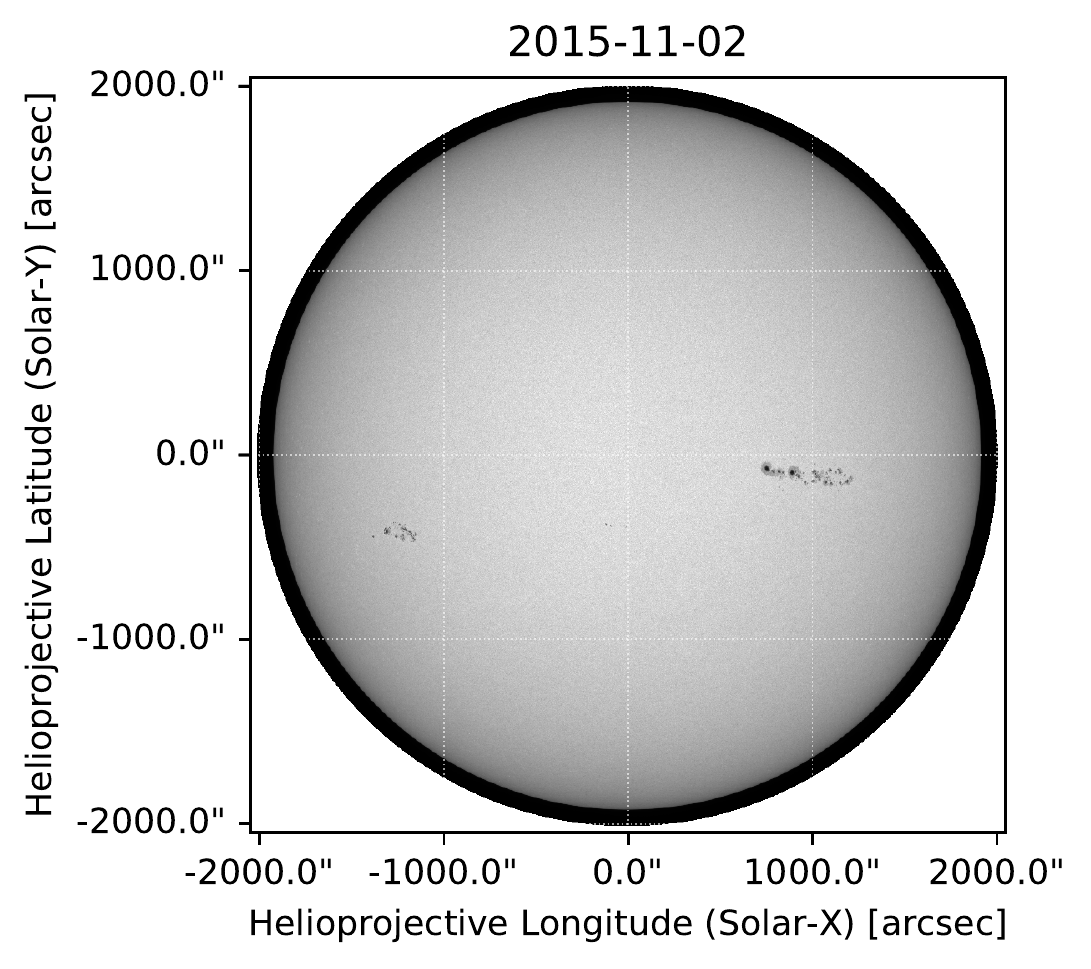}
    \caption{\textit{Left:} cross-correlation functions of the 4000 K TiO template emission spectrum and two HARPS spectra of the Sun (reflected off of the Moon) on November 2, 2015 -- a night with two large sunspot groups facing the Earth. The echelle order used in these plots is the one centered on 6797 \angstrom, which is the order with the strongest CCF signal when cross-correlating the Proxima Centauri spectrum with the 3000 K TiO template. The vertical dashed line is the estimated radial velocity of the source. \textit{Right:} SDO HMI continuum intensity image of the Sun during the HARPS observations, revealing the large sunspot groups which are not detected by the cross-correlation technique.}
    \label{fig:sun}
\end{figure*}

For several thousand publicly available lunar spectra from HARPS, we retrieve simultaneous Solar Dynamics Observatory (SDO) HMI continuum intensity imaging of the Sun. We find that the Sun was most spotted during lunar HARPS observations on UTC November 2, 2015 (Program ID: 096.C-0210(A), PI: P.~Figueira), when two major spot groups were on the Earth-facing solar hemisphere -- see Figure~\ref{fig:sun}. 

We cross correlate the solar spectrum with the 4000 K TiO template. We find no significant absorption in any of the nine observations which have $S/N > 100$ on that night, in any of the four echelle orders where the TiO line list is expected to produce the strongest CCF signal based on the cross-correlation with the spectrum of Proxima Centauri. 

\subsection{Search for starspots on 18 Sco} \label{sec:18sco}

For more than 20 years, 18 Scorpii has been studied as a solar twin \citep{Mello1997}, that is, a star with spectroscopic parameters exceptionally similar to the Sun's \citep{CayreldeStrobel1996}. \citet{Hall2000} and \citet{Hall2007} showed that 18 Sco has a seven year activity cycle that is similar to the Sun's in terms of total irradiance variation. Joint asteroseismic and spectroscopic analyses have yielded highly precise measurements of the stellar radius and mass \citep{Bazot2011,Li2012,Bazot2012}. \citet{Petit2008} used spectropolarimetry to show that its rotation period is $22.7 \pm 0.5$ days, only a few days shorter than solar, and more recently \citet{Bazot2018} used asteroseismology to suggest that the age of 18 Sco may be consistent with solar (though estimates have varied from 0.3 Gyr to 5.8 Gyr, \citealt{Tsantaki2013, Takeda2008, Mittag2016}).  Even under intense scrutiny, this star continues to appear remarkably similar to the Sun, so one might expect 18 Sco to have spots like the Sun does.

18 Sco has been the subject of various radial velocity searches for exoplanets with HARPS. We gather 4000 archival HARPS spectra of 18 Sco collected since 2006 under various observing programs\footnote{Program IDs: 188.C-0265(R), 072.C-0488(E), 188.C-0265(L), 188.C-0265(E), 185.D-0056(E), 183.D-0729(A), 198.C-0836(A), 183.D-0729(B), 188.C-0265(P), 188.C-0265(O), 0100.D-0444(A), 188.C-0265(C), 196.C-1006(A), 188.C-0265(Q), 188.C-0265(K), 188.C-0265(G), 183.C-0972(A), 192.C-0852(A), 188.C-0265(J), 188.C-0265(H), 099.C-0491(A), 188.C-0265(D)}. We stack all spectra of 18 Sco together by shifting the wavelength axis of each spectrum to maximize the cross-correlation with the previous spectrum. The coadded spectrum has $S/N\approx2200$. 

\begin{figure*}
    \centering
    \includegraphics{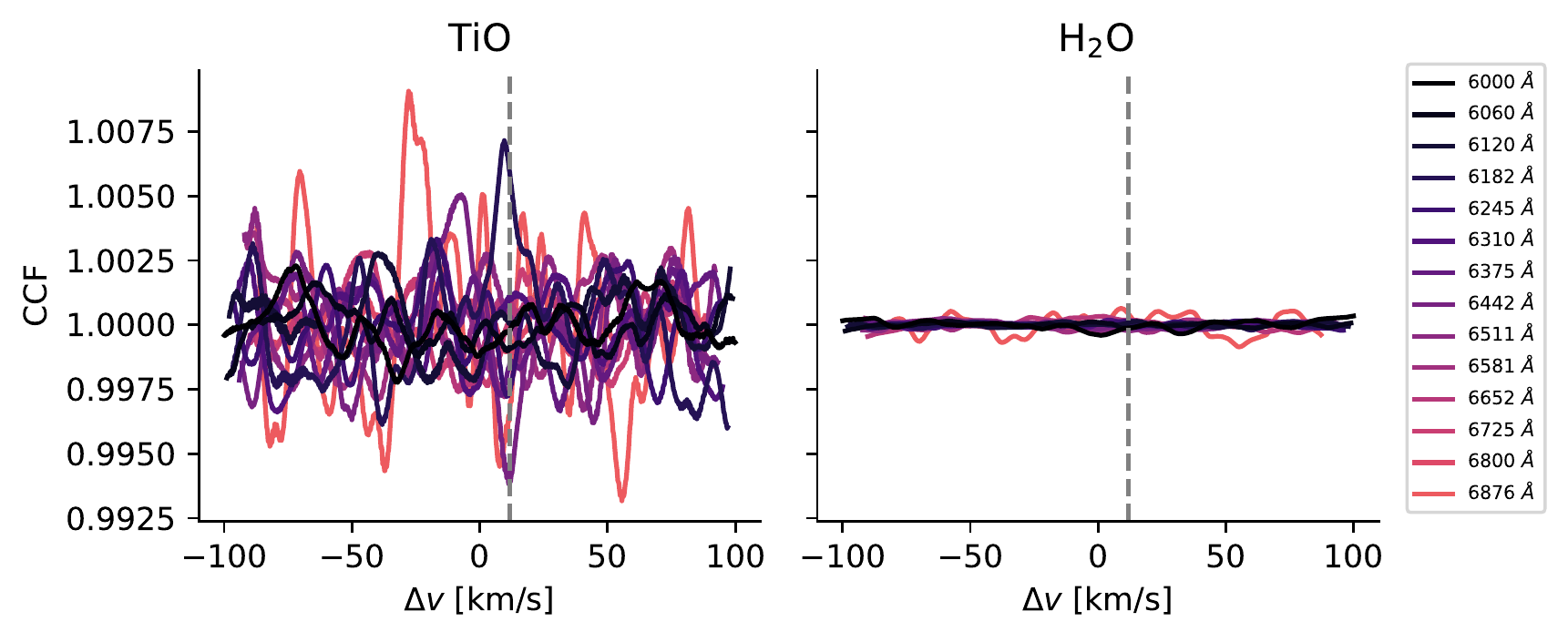}
    \caption{Cross-correlation function of 4000 stacked HARPS spectra of 18 Scorpii with the TiO ($\tspot=4000$ K) and water ($\tspot = 2500$ K) templates, with a coadded $S/N\approx2200$. The CCF produces no signal with either molecule, indicating a small spot coverage for 18 Sco. The vertical dashed line is the estimated radial velocity of the source.}
    \label{fig:18sco}
\end{figure*}

The CCF of the stacked HARPS spectra and the TiO and water emission templates are shown in Figure~\ref{fig:18sco}. Each curve represents the CCF of a single echelle order with the template. If TiO or water absorption were present in the coadded spectrum, there would be a negative absorption feature dipping below unity near the radial velocity of the star ($v=11.90$ km s$^{-1}$), but no signal is detected. 

In the case of the spectrum of 18 Sco, we have a Sun-like star with an unknown spot coverage and a coadded $S/N\approx2200$. The null detection of water and TiO in the stacked spectrum of 18 Sco given the simulations in Section~\ref{sec:sims} places an upper limit on the spot covering fraction $f_S \lesssim 4 \times 10^{-3}$. This is smaller than the Sun's most extreme spot coverages near solar maximum, $f_S \approx 5 \times 10^{-3}$ \citep{Morris2017a}.

\subsection{Search for the spots of HAT-P-11}

HAT-P-11 is an active K4V dwarf in the \kepler field with a transiting hot Neptune. Transits revealed frequent starspot occultations \citep{Deming2011,Sanchis-Ojeda2011}, which yield approximate spot covering fractions $f_S = 0.0-0.1$ \citep{Morris2017a}. HAT-P-11 appears to have a $\sim$10 year activity cycle, and may be modestly more chromospherically active than planet hosts of similar rotation periods \citep{Morris2018b}. Recent ground-based photometry of spot occultations within the transit chord yielded spot coverage $f_S = 0.14$ \citep{Morris2018d}. \citet{Morris2019b} model the spectrum of HAT-P-11 as a linear combination of the spectra of HD 5857 and Gl 705, giving the spots $\Delta T_\mathrm{eff} = 250$ K, similar to typical sunspot penumbra, finding a spot coverage consistent with previous measurements. 

\begin{figure}
    \centering
    \includegraphics[scale=0.8]{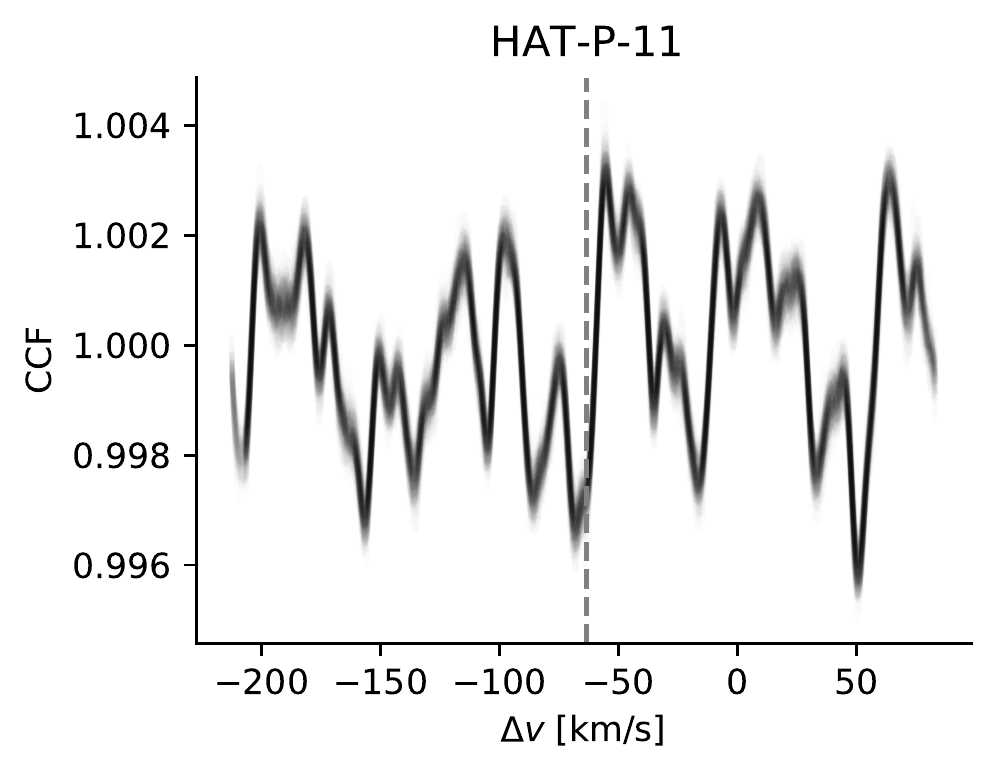}
    \caption{Cross-correlation functions of the HARPS-N spectra of the highly-spotted K4V star HAT-P-11 and the TiO template with $\tspot = 4000$ K. The echelle order used in these plots is the one centered on 6722 \angstrom, which is the order with the second strongest CCF signal when cross-correlating the Proxima Centauri spectrum with the 3000 K TiO template. The vertical dashed line is the radial velocity of the star.}
    \label{fig:hatp11}
\end{figure}

We cross-correlate 139 HARPS-N spectra of HAT-P-11 (Program ID: OPT15B\_19, PI: D.~Ehrenreich) with the TiO template at 4000 K in Figure~\ref{fig:hatp11}. There is no significant absorption in the CCF due to TiO, despite HAT-P-11 being on average 100x more spotted than the Sun \citep{Morris2017a}.

\subsection{Search for the cool regions of T Tauri Stars: LkCa 4 and AA Tau}

LkCa 4 is a T Tauri star which is often classified as a K7 dwarf \citep{Herbig1986}. \citet{Gully2017} used high-resolution near-infrared IGRINS spectra to show that the stellar surface of LkCa 4 is in fact dominated by cool regions, covering 80\% of the stellar surface with $T_\mathrm{cool} \sim 2700-3000$ K. Hot regions make up the other 20\% of the surface with $T_\mathrm{hot} \sim 4100$ K. 

\begin{figure*}
    \centering
    \includegraphics[scale=0.8]{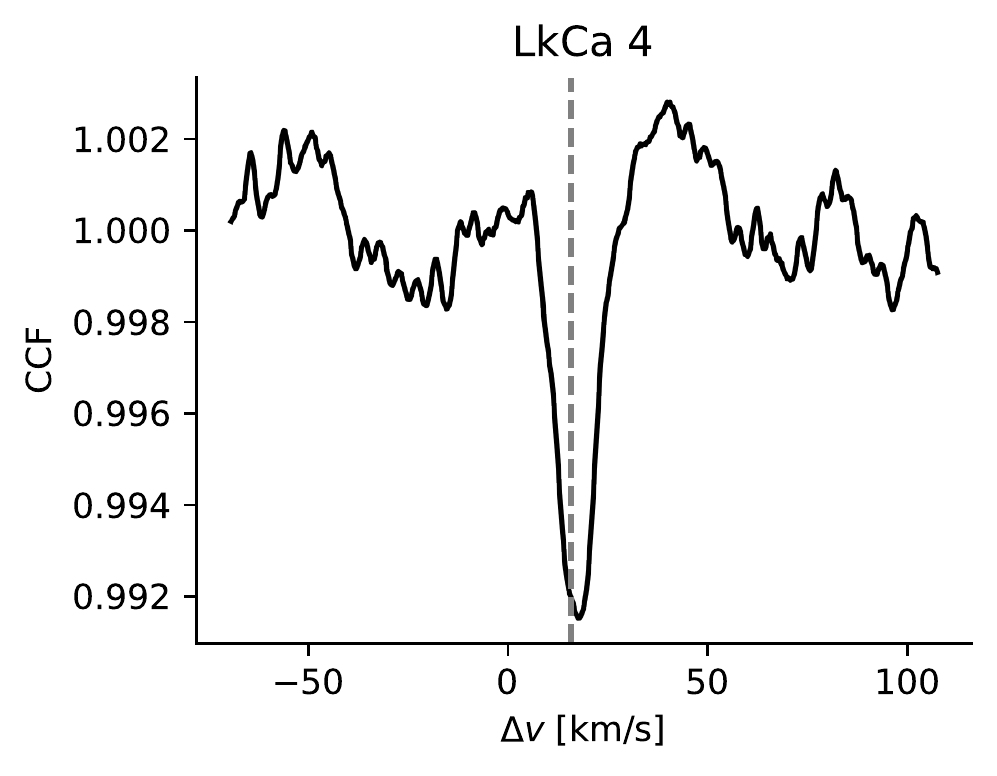}
    \includegraphics[scale=0.8]{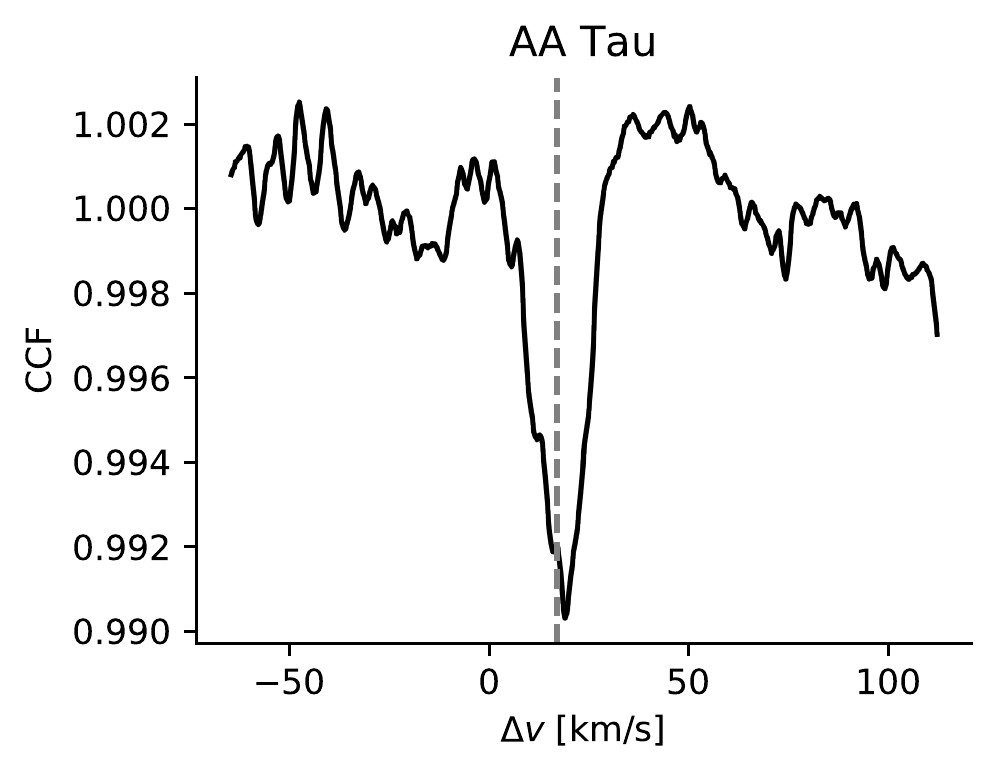}
    \caption{Cross-correlation functions of the HARPS spectra of the puzzling K7 dwarf LkCa 4 (left) and the K5 dwarf AA Tau (right), and the TiO template with $\tspot = 4000$ K. The echelle order used in these plots is the one centered on 6800 \angstrom, which is the order with the strongest CCF signal when cross-correlating the Proxima Centauri spectrum with the 3000 K TiO template. The vertical dashed line is the radial velocity of each star.}
    \label{fig:lkca4}
\end{figure*}

We examine the CCF of the LkCa 4 HARPS spectrum (Program ID: 074.C-0221(A), PI J.~Bouvier) as a control to verify that TiO can be detected in stars earlier than M-type, when they are known to be extremely ``spotted.'' Figure~\ref{fig:lkca4} shows the CCF, confirming strong absorption features due to TiO near $\tspot = 4000$ K. The clear CCF signal confirms that indeed the star has significant coverage by regions cooler than the K7 spectral type assigned to this star.

AA Tau is a K5 dwarf, and also a T Tauri star. We cross-correlate the HARPS spectrum of AA Tau (Program ID: 074.C-0221(A), PI J.~Bouvier) with the TiO emission template at $\tspot=4000$ K. Again, we find evidence for significant absorption by TiO in the atmosphere of this ``K'' star, confirming that at least one other T Tauri star has significant coverage by cool regions, and that our CCF technique is performing as expected on a highly-spotted control star.

\section{Conclusion} \label{sec:conclusion}

Starspots on Sun-like stars are functionally invisible in HARPS/HARPS-N spectra when using TiO as a tracer. While the invisibility of starspots to the CCF technique may dismay starspot hunters, exoplanet hunters searching for molecular absorption in exoplanet atmospheres can be confident that the signals they detect come from the exoplanet rather than the star. Starspots should be an insignificant source of TiO absorption in the spectra of exoplanetary systems with FGK host stars. Of course, one could also disentangle the stellar and planetary signals with to the difference in velocity between the star and the planet.

\acknowledgements
           
This work has been carried out in the framework of the PlanetS National Centre of Competence in Research (NCCR) supported by the Swiss National Science Foundation (SNSF). This research has made use of the VizieR catalogue access tool, CDS, Strasbourg, France. The original description of the VizieR service was published in A\&AS 143, 23. This research has made use of NASA's Astrophysics Data System. Based on observations made with ESO Telescopes at the La Silla Paranal Observatory. Based on observations made with the Italian Telescopio Nazionale Galileo (TNG) operated on the island of La Palma by the Fundaci\'{o}n Galileo Galilei of the INAF (Istituto Nazionale di Astrofisica) at the Spanish Observatorio del Roque de los Muchachos of the Instituto de Astrofisica de Canarias.

\software{\texttt{astropy} \citep{Astropy2013, Astropy2018}, \texttt{sunpy} \citep{Sunpy2015}, \texttt{ipython} \citep{ipython}, \texttt{numpy} \citep{VanDerWalt2011}, \texttt{scipy} \citep{scipy}, \texttt{matplotlib} \citep{matplotlib}, \texttt{FastChem} \citep{Stock2018MNRAS.479..865S}, \texttt{CDISORT} \citep{Hamre2013AIPC.1531..923H}, \texttt{Helios-o} \citep{Bower2019}}

\facilities{HARPS, HARPS-N}

\appendix

\section{Spectral Templates}

\begin{figure*}[h]
    \centering
    \includegraphics[scale=0.9]{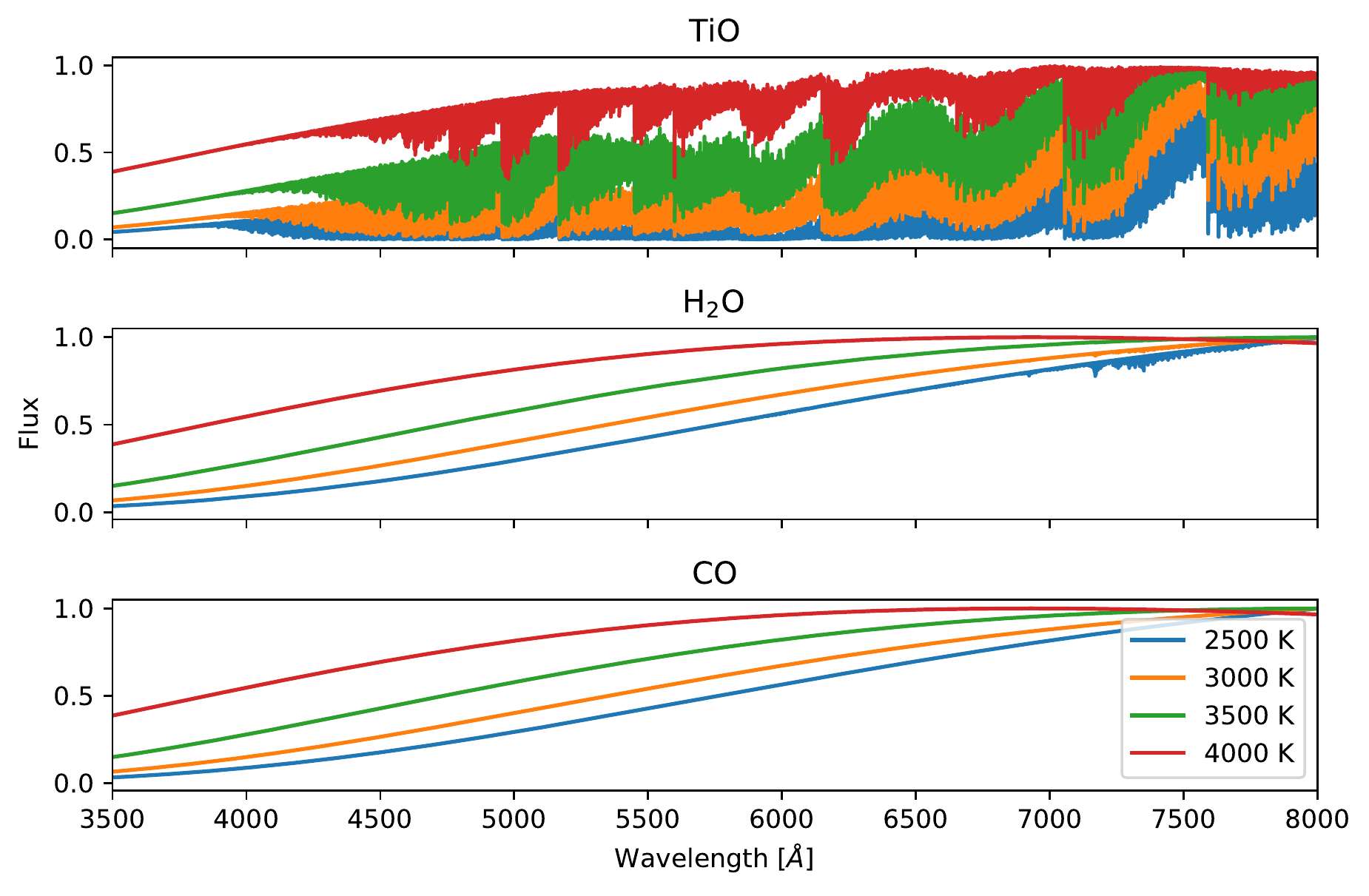}
    \caption{Spectral templates for TiO, H$_2$O, and CO at four temperatures. TiO shows significant absorption bands throughout the HARPS bandpass at all temperatures, while water only deviates significantly from a blackbody at $T < 3000$ K. CO never deviates significantly from a blackbody in the temperature range and HARPS bandpass.}
    \label{fig:molecules}
\end{figure*}

Figure~\ref{fig:molecules} shows the resulting TiO spectral templates each normalized by their maximum flux.

\end{document}